# Evolution of the genetic code. Why are there strong and weak letter doublets? The first gene, the first protein. Early (ancient) biosynthesis of protein.

Semenov D.A. (dasem@mail.ru)

International Research Center for Studies of Extreme States of the Organism at the Presidium of the Krasnoyarsk Research Center, Siberian Branch of the Russian Academy of Sciences

*The idea of the evolution of the genetic code from the CG to the CGUA alphabet has been developed further. The assumption of the originally triplet structure of the genetic code has been substantiated. The hypothesis of the emergence of stop codons at the early stage of the evolution of the genetic code has been additionally supported. The existence of strong and weak letter doublets of codons and the symmetry of strong doublets in the genetic code table, discovered by Rumer, have been explained. A hypothesis concerning the primary structure of the first gene and the first protein has been proposed.*

**Strong and weak letter doublets. Rumer's symmetry.**

Rumer proposed this representation of the table of letter doublets for the genetic code in order to illustrate the symmetry he had discovered [1,2] (see Table 1).

|   | **C** | **G** | **U** | **A** |
|---|---|---|---|---|
| **C** | Pro | Arg | Leu | His<br>Gln |
| **G** | Ala | Gly | Val | Asp<br>Glu |
| **U** | Ser | Cys<br>Trp/Stop | Phe<br>Leu | Tyr<br>Stop |
| **A** | Thr | Ser<br>Arg | Ile<br>Met | Asn<br>Lys |

Table 1. The symmetry of the table of the letter doublets for the genetic code (according to Rumer). Strong letter doublets are marked in gray.

A nucleotide triplet can be assumed to change its form, but in this case a change in the form of the first two nucleotides occurs due to the third. Here I do not speak of the form of the triplet in the solution or the form of the triplet in mRNA, but rather the form of the double helix segment resulting from the codon-anticodon interaction.

This form is determined by two forces – complementary interactions and stacking (interaction of neighboring nucleotides). Stacking is a nonspecific event as there is a well-known formula: purine-purine>purine-pyrimidine>pyrimidine-pyrimidine.

In the third position of the codon the number of the formed hydrogen bonds is much less relevant than the nature of the bases – purine or pyrimidine. Thus, the presence of strong and weak letter doublets can naturally be related to the presence of stacking.

In codons with the CC, CG, GC, and GG letter doublets, the form of the doublet is only determined by complementary interaction. The three hydrogen bonds in each doublet make conformational changes impossible.

Let us now discuss the (UC-UG), (AC-AG), (CU-CA) and (GU-GA) letter doublets. In each doublet the first letter is strong and the second weak. The number of hydrogen bonds in each doublet is the same, but doublets with purine in the second position are the weak ones. Due to the presence of purine in the second position, two purines can occur one by one – in the second and the third positions of the codon. This construction imparts sufficient stress to the first two nucleotides to cause a change in their conformation. Please note that the anticodon is much less conformationally flexible because it is part of tRNA and is stabilized by its structure.

In codons with the UA and AA doublets, the conformation of the doublet can be changed due to the presence of purine in the third position of the codon. In the codons with the UU and AU doublets, the conformation of the codon can be changed by just a slight interaction between pyrimidine and purine.

The fact that complementary interactions are comparable with stacking in their effect may be surprising, but if this were not so and there were just one predominant type of nucleotide interactions, the other would not be described in textbooks. It is not less surprising that only half of the codon letter doublets are capable of conformational changes under the impact of stacking. This may be a significant fact, but I have no ideas concerning this.

Rumer's symmetry can then be interpreted as follows: U>A because pyrimidine in the second letter of the codon prevents it from possible stacking-related conformational changes. This is not related to the sequence in which nucleotides were incorporated into the code. Crick's sequence U>C>A>G can serve to illustrate similarities inside the class of purines and pyrimidines and the possibility of replacing G by U in the first letter. The A>G>C>U sequence has the same properties. The table of the genetic code presented in the circular form demonstrates these properties even better. The circular form of the table is the best representation of the symmetry, which ensues from evolution. Thus, I was too hasty to draw the conclusion that the U>A and C>G relationships were universal and evolutionarily significant.

In the table of the universal genetic code there are two amino acids, each of which is encoded by one codon only: tryptophan (Trp), encoded by the UGG codon, and methionine (Met), encoded by AUG. This feature can be consistently explained based upon the same arguments that have been used to explain Rumer's symmetry. If there are letter doublets incapable of conformational changes when pyrimidine is replaced by purine in the last letter and there are letter doublets that readily change under these conditions, there must be letter doublets that are close to the equilibrium point. For these letter doublets even much smaller impacts may prove to be significant. The UG doublet is situated on the table diagonal, i.e. it may be close to the equilibrium point. For its conformational change it may be important not only that the third position is occupied by purine but also that this is guanine. An additional hydrogen bond leads to a conformational change. The situation with the AU doublet is similar, but instead of the purine-purine interaction, we have to assume that its conformation can be changed by a weaker, pyrimidine-purine, interaction. In this case, the letter doublet itself is extremely prone to conformational changes: the doublet contains no nucleotides capable of forming three hydrogen bonds. The AU doublet is presumably close to the equilibrium point and the conformational transition is achieved at the expense of one hydrogen bond in the last nucleotide.

Thus, it is the first two nucleotides, or, to be more exact, the conformation of the first two nucleotides, that are the encoding part of the codon. For half of the letter doublets this conformation depends upon the last nucleotide, although it is this conformation rather than the whole sequence that is recognized. The number of different letter doublets of codons is not large: 8 (one for each strong doublet) + 16 (two for each weak one) = 24. There is only one codon for which all three nucleotides are significant – UGA, the stop codon. UGA's potential anticodon – UCA, placed into the tryptophan tRNA, would not be able to encode tryptophan. It would keep the conformation of the first two nucleotides unchanged. This can be a way to verify my speculations experimentally.

**The triplet genetic code.** Speaking of the codon size, I suppose that it was originally triplet, even at the stage when there were just guanine and cytosine. In my opinion, this is not related to the necessity for four nucleotides to encode for twenty amino acids; it is rather the number of amino acids that is limited by the number of encoding variants of codon forms. Assuming that the code expanded incorporating new amino acids, it would be more natural to suggest that the number of amino acids should be limited by the capacity of the code rather than that the length of the encoding word should be determined by the number of amino acids.

The triplets represented by guanine and cytosine only cannot encode for more than four amino acids. This is in good agreement with my substantiation of Rumer's symmetry and the above suggestion that the form of the letter doublet is the recognizable component of the codon.

There is no need to speak of the doublet code, automatically applying Gamov's ideas to the early genetic code [3]. A simpler assumption is that the length of the codon has always been constant, and just its composition has been changed.

This assumption allows a natural explanation of why the replacement of purine by pyrimidine is extremely significant in the second letter and is often permissible in the first letter of the codon. Changes in the first letter of the triplet alter the shape of the doublet (the size of the first letter is altered), but this alteration affects the surroundings of the middle letter only. Changes in the second letter alter the surroundings of all letters, the whole shape of the codon. For the doublet code, changes would be equally significant, and insignificant variations in kindred codons would be out of the question. This does not give rise to any contradictions in the stage when there were just C, G, and U, because the only mutation that I dealt with in that stage [4] was cytosine deamination. Cytosine deamination does not alter the size of the nucleotide. In that stage, however, I expect the emergence of the stop codon, and this can be more easily accounted for based on the triplet code hypothesis, as will be clear from the section below.

**Emergence of the stop codon.** I have already put forward the hypothesis that the stop codon emerged as a result of instability of interaction between the UG doublet and its potential anticodon [5]. For this hypothesis it is essential to recognize the existence of the stage at which there is uracil but there is no adenine yet. As stated above, in this stage uracil is incorporated in the RNA double helix in its enol form. This is also valid for the codon-anticodon interaction, but specific properties of the micro-surroundings cannot be ignored. I suppose that the codon beginning with the UG doublet had originally been prone to have uracil in the keto form in its complementary interaction with the potential anticodon, which, respectively, began with GC.

What could be the difference between this codon and the other codons? There are grounds to suppose that its properties are related to the symmetry of strong and weak letter doublets discovered by Rumer. This was the only weak doublet that had emerged before adenine did.

Let us first look at the CU and GU doublets. If we assume the existence of the doublet code, uracil is at the edge and can be in its keto form. If the code is triplet, the small portion of uracil in the RNA of that stage makes G or C the most probable third letter. The enol form of uracil, restricted in this way, will be able to make at least two hydrogen bonds with the complementary guanine. This may be interpreted as an indirect proof of the code being triplet, but this indirect proof can be verified experimentally!

Let us now compare triplet codons with the UC and UG doublets. The UG doublet does not change its conformation in the contemporary genetic code and it seems that it was not prone to do this before. There is no evidence of uracil preferably being in the enol form, but, at the same time, nothing points to a greater stability of the keto form in the absence of adenine.

The UG doublet is very illustrative: it is both able to change its conformation and sensitive to impacts equal to the energy of one hydrogen bond. In this case, one hydrogen bond can come into being and cease to exist when uracil is transformed from the enol form to the keto one. To change the conformation, purine does not have to be replaced by pyrimidine in the third letter. The same force that was earlier (in discussing the tryptophan codon) applied at one side is now taken away at the other. The transformation of uracil into its keto form is stimulated by the inclination of this (triplet) codon towards conformational changes. This reasoning is certainly less forcible than experiment (quite feasible in this case), but more convincing than many of quantum chemical calculations. Please note again that for this reasoning it is essential that the triplet code should have existed even when there were just C, G, and U.

If the CU, UC, and GU doublets are characterized by the presence of uracil in the enol form and the UG doublet – by the presence of uracil in the keto form, the loss of one (or even two) hydrogen bonds in the interaction with the respective anticodon can lead to the development of stop codon properties in a "natural" way. Again, these properties can be studied experimentally, in contrast to discussions concerning evolution of aminoacyl-tRNA synthetases.

**The first gene, the first protein.** With just the first four amino acids encoded by guanine and cytosine there is no hope for the synthesis of catalytically active protein. Proline, glycine, alanine, and arginine – no combination of these amino acids can yield a wide diversity of structures and properties.

Moreover, I assume that there was no start signal or translation stop signal. One can only suppose that synthesis could start at any position. This assumption makes it equally probable for any codon to occur. Supposedly, the primary sequence of early genes was random, which would suggest the stability towards the reading-frame shift. This assumption cannot be regarded as sufficiently productive, as such a starting point of evolution is extremely far away from any sequence of selective value.

Let us imagine the maximally ordered structure of the first gene, which is similar to the crystal structure. This structure would be advantageous for self-reproduction. The existing disordering factors (spontaneous mutations) will in the future cause the periodic crystal to be transformed to Schrödinger's "aperiodic" crystal [6].

The fact that it is equally probable to find any of these four amino acids in the protein suggests one assumption concerning the properties and structure of the first protein. The first protein could be either collagen or collagen-like protein. High proline content is a necessary condition for constructing collagen. It is the presence of proline in every third or fourth position that makes the formation of the triple helix possible.

An essential condition for collagen is strict periodicity of the primary structure. The triplet code and strict periodicity make the structure sensitive to the reading-frame shift. If the task is to determine the structure of the gene stable towards any reading-frame shift, this problem, no matter how surprising this may seem, has one rigorous solution:

(CCGG)(CCGG)(CCGG)(CCGG)(CCGG)(CCGG) etc.

The primary protein sequence will consist of the repeated motif (ProAlaGlyArg).

This gene and this protein allow fulfilling the invariance condition for transposition of RNA strands and invariance to the reading-frame shift; moreover, this is the only solution.

Could there be any reasons for realizing this very gene in the form of the RNA double helix? Let us consider the extreme case of the RNA double helix, in which one strand is represented by polyguanine and the complementary one – by polycytosine. In their free states, in the solution, polycytosine and polyguanine can be considerably different from each other. The strong influence of stacking in polyguanine will result in the structure with a significantly extended helical turn. The formation of the double helix will be associated with an essential restructuring of the helix. With the polymer being sufficiently long, the kinetic barrier can be unsurpassable. The symmetric positions of cytosine and guanine in the complementary strands make their forms similar and, thus, not requiring large expenditures of energy and time for the formation of the double helix. Simple recurrence of the GC fragment is disadvantageous, because in this structure the influence of stacking is very low. The repeated motif (CCGG) may on the average lead to some energy benefit. To continue the logic of this reasoning, one can assume that the (CCCGGG) structure is even more stable, etc. As already stated in this section, polyguanine and polycytosine must have incompatible structures of RNA single helixes. What should be the length of oligomer to reveal this incompatibility? My substantiation of Rumer's symmetry of the genetic code table suggests that stacking considerably influences the shape of the RNA double helix segment formed by the codon and the anticodon. That is, even within one triplet, the length of the helical turn can be noticeably different in polycytosine and polyguanine. This difference is noticeable for the formation of the double helix.

In the proposed structure (CCGG) these differences are physically (or mathematically) impossible.

The idea that the evolution of protein biosynthesis started with a strictly periodic gene rather than with quasi-periodic structures (as proposed by Crick) can certainly be received skeptically. However, this initial state, perfectly stable towards the reading-frame shift has some advantages. There are genes that simultaneously encode for two proteins, as a result of the reading-frame shift. The existence of these genes contradicts the assumption of the initial existence of only such genes as are written without a reading-frame shift. Based on the evolution of all genes from the single initial state, which is perfectly stable to the reading-frame shift, one can study the models of the evolutionary emergence of genes encoding for more than one protein with a shift.

Another issue that may be approached in a way that is "automatically" determined by the presence of the poly(CCGG) structure as the first gene is the emergence of tRNA. The primary structure of tRNA must contain several complementary segments. The poly(CCGG) gene has the necessary properties intrinsically. The single-strand molecule of this RNA must form the secondary structure through the formation of double helix segments. An approach to the emergence of ribosomal RNAs can also be proposed: their secondary structure is also predetermined by the availability of significant complementary segments in the primary sequence. The existence of this first gene will imply the existence of the structures similar to the tRNA and ribosomal RNA even prior to their acquisition of their contemporary functions.

**Early biosynthesis of protein.**

Once I have made assumptions concerning the early genetic code and even the first gene and the first protein, I should propose a possible mechanism of protein biosynthesis. In my opinion, the most likely mechanism is direct biosynthesis of protein on the polynucleotide matrix, already proposed by Gamov in 1953 [3], in the first study of the genetic code.

Note that arginine has been demonstrated to be able to recognize its codon. The other three earliest amino acids do not carry a positive charge and cannot be sorbed by the RNA polyanion so well. I can propose changing experimental conditions by supplementing the medium with metal cations, which can form complexes with RNA and amino acids. Magnesium and calcium cations can be the most suitable as they are ubiquitous. This can be a way to equalize the charge.

The above arguments suggest that the dinucleotide form is the meaningful part of the codon and this is valid for the whole table of the genetic code. This can be considered as an "atavism" of the epoch in which amino acids were directly recognized by codons. Then, what is the function of the third nucleotide? Although it did not take part in the recognition, it was most probably necessary for the amino acid-codon geometric correspondence – just a dimensional correspondence between the amino acid and the codon. Thus, the existence of the triplet code in the early phases of the evolution of the genetic code contradicted the logic of Gamov's study [3]. Yet, we should note that in the hypothetical first gene there is just one codon corresponding to each amino acid.

Gamov assumed that protein was synthesized immediately on the DNA double helix [3]. My assumption is immediate protein synthesis on the RNA double helix. Segments of the RNA double helix still work in ribosomes. Unlike DNA, the RNA double helix is rigid enough to set considerable segments into coordinated motion. This is very useful for catalysis.

Let us now address the mechanism of catalysis. Let us look at the RNA (DNA) double helix close to the "melting" point, where the double helix is untwined to make two single-strand molecules. This process is accompanied by a disruption of three hydrogen bonds in one base pair and binding of six molecules of water. This is a mightiest dehydration effect.

The reaction of the formation of the peptide bond during protein synthesis is essentially dehydration reaction. Two amino acids lose one water molecule.

If amino acids have affinity for their codon, the formation of these complexes must induce melting of the double helix and subsequent dehydration. If two amino acids are joined to RNA next to each other, one water molecule can leave them and a peptide bond can be formed. Partial untwining of the double helix makes new RNA segments accessible and they are joined by new amino acids; thus, protein biosynthesis goes on.

The newly formed protein must be a mechanical obstacle to the restoration of the disrupted segment of the double helix and cannot be recognized by hydrogen bonds of nucleotides any more.

This mechanism makes biosynthesis possible without the specific catalyst moving along the RNA. It offers a probable dehydrating reagent. It rules out adaptors and reduces protein biosynthesis to the minimal number of interacting reagents.

Beginning with this state, the further evolution is potentially deducible.

Please note that guanine and cytosine have a number of advantages for this scheme: 1. The ability to form the RNA double helix. 2. The presence of a large number of hydrogen bonds for molecular recognition. 3. The availability of the best opportunities for binding water molecules. These advantages are all based on the possibility of forming three complementary hydrogen bonds.

**Acknowledgement**
The author would like to thank Krasova E. for her assistance in preparing this manuscript.

**References:**
1. Rumer Yu.B. On systematizing codons in the genetic code. *Doklady Akademii Nauk SSSR* 167, 6, 1394, (1966).

2. Rumer Yu.B. Systematizing of codons in the genetic code. *DAN SSSR.* 183, 1, 225-226, (1968).
3. Gamow G. Possible Relation between Deoxyribonucleic Acid and Protein Structures, Nature Vol. 173, 318, (1954).
4. Semenov D.A. Evolution of the genetic code from the GC- to the AGUC-alphabet. arXiv.0710.0445
5. Semenov D.A. Evolution of the genetic code. Emergence of stop codons. arXiv.org.0710.5825
6. Schrödinger E. What is Life? Macmillan (1946).